\documentclass[pra,aps,amsmath,amssymb,showpacs,twocolumn]{revtex4}

\usepackage{graphicx}
\usepackage{amsmath}
\newcommand{\Hil}{{\mathcal H}}

\newcommand{\Boltz}{ k_{\rm\scriptscriptstyle B}}
\newcommand{\Tr}{{\rm Tr}}
\newcommand{\Ran}{{\rm Ran}}

\newcommand{\citen}[1]{{ \cite{#1}}}

\begin{document}

\title{The Hatsopoulos-Gyftopoulos  resolution of the Schr\"odinger-Park
paradox about the concept of ``state'' in quantum statistical
mechanics}
\author{Gian Paolo Beretta}
\affiliation{ Universit\`a di Brescia, via Branze 38, 25123
Brescia, Italy } \email{ beretta@unibs.it}
\date{\today}
%


\begin{abstract}
A seldom recognized fundamental difficulty undermines the concept
of   individual ``state'' in the present formulations of quantum
statistical mechanics (and in its quantum information theory
interpretation as well). The difficulty is an unavoidable
consequence of an almost forgotten corollary proved by E.
Schr\"odinger in 1936 and perused by J.L. Park, Am. J. Phys. {\bf
36}, 211 (1968). To resolve it, we must either reject as unsound
the concept of state, or else undertake a serious reformulation of
quantum theory and the role of statistics. We restate the
difficulty and discuss a possible resolution proposed in 1976 by
G.N. Hatsopoulos and E.P. Gyftopoulos, Found. Phys. {\bf 6}, 15,
127, 439, 561 (1976).
\end{abstract}

%

\maketitle

\section{Introduction}

In 1936, Schr\"odinger \cite{Schroedinger} published an article to
denounce a ``repugnant" but unavoidable consequence of the present
formulation of Quantum Mechanics (QM) and Quantum   Statistical
Mechanics (QSM). Schr\"odinger claimed no priority on the
mathematical result, and properly acknowledged that it is hardly more
than a corollary of a theorem about statistical operators that von
Neumann proved  five years earlier \cite{Neumann}.

Thirty years later, Park \cite{Park1} exploited von Neumann's
theorem and Schr\"odinger's corollary to point out quite
conclusively an essential tension undermining the logical
conceptual framework of QSM (and of its Quantum Information Theory
interpretation as well). Twenty more years later, Park returned on
the subject in another magistral, but almost forgotten paper
\cite{Park2} in which he addresses the question of ``whether an
observer making measurements upon systems from a canonical
ensemble can determine whether the systems were prepared by
mixing, equilibration, or selection'', and concludes that ``a
generalized quantal law of motion designed for compatibility with
fundamental thermodynamic principles, would provide also a means
for resolving paradoxes associated with the characteristic
ambiguity of ensembles in quantum mechanics.''

Schr\"odinger's corollary was ``rediscovered'' by Jaynes
\cite{Jaynes1,Jaynes2} and Gisin \cite{Gisin}, and  generalized by
Hughston, Jozsa, and Wooters \cite{Hughston} and Kirkpatrick
\cite{Kirkpatrick}. Also some interpretation has been
re-elaborated around it
\cite{Mermin,Amann,Wiseman,Cohen1,Cohen2,Terno}, but unfortunately
the original references have not always been duly cited. The
problem at issue in this paper, first raised
 in Ref.\ \citen{Schroedinger}, has
been acknowledged ``in passing'' in innumerable other references
(see, e.g., Refs. \cite{Elsasser,Allahverdyan} and references
therein), but none has to our knowledge gone so deeply and
conclusively to the conceptual roots as Refs.\
\citen{Park1,Park2}. For this reason it is useful once in a while
to refresh our memory about the pioneering conceptual
contributions by Schr\"odinger and Park. The crystal clear logic
of their analyses should not be forgotten, especially if we decide
that it is necessary to ``go beyond''.  Ref.\ \citen{Schroedinger}
has been cited by many others, but not about the problem we focus
on here, rather for it also contains pioneering contributions to
the question of entanglement, EPR paradox and related nonlocal
issues. Both Refs.\ \citen{Schroedinger} and \citen{Park1,Park2}
have been often cited also in relation to the projection postulate
and the quantum measurement problem.

 The tension that Park vividly brings out in his beautiful
  essay on the ``nature of quantum states''
  is about the central
concept   of individual state of a system. The present formulation
of QM and the standard interpretation of QSM imply the paradoxical
conclusion that every system is ``a quantum monster'': a single
system can be thought as concurrently being ``in" two (and
actually even more) different states. We briefly review the issue
below (as we have done also in Ref.~\citen{thesis}), but we urge
everyone interested in the foundations of quantum theory to read
the original references \cite{Park1,Park2}. The problem has been
widely overlooked and is certainly not well known, in spite of the
periodic rediscoveries. The overwhelming successes of QM and QSM
understandably contributed to discourage or dismiss as useless any
serious attempt to resolve the nevertheless unavoidable
fundamental difficulty.

Here, we emphasize that a resolution of the tension requires a
serious re-examination of the conceptual and mathematical
foundations of quantum theory.  We discuss three logical
alternatives. We point out that   one of these alternatives
achieves a resolution of the fundamental difficulty without
contradicting any of the successes of the present mathematical
formalism in the equilibrium realm where it is backed by
experiments. However, it requires an essentially new and different
re-interpretation of the physical meaning of such successes.
Moreover, in the nonequilibrium domain it opens to new
discoveries, new physics compatible with the second law of
thermodynamics, without contradicting QM, and resolving the
Boltzmann paradox about irreversibility as well. Thermodynamics
may thus play once again  a key role in a conceptual advancement
\cite{HG1,HG2,HG3,HG4,NATO,Steepest,JMP1,Cimento1,Cimento2} which
may prelude to uncovering new physics about far non-equilibrium
dynamics \cite{Gheorghiu1,Gheorghiu2,ArXiv1,MPLA,PRE}.

\section{Schr\"odinger-Park quantum monsters}

In this section, we review briefly the problem at issue.  We start
with the seemingly harmless assumption that every system is always
in some definite, though perhaps unknown, state. We will conclude
that the assumption is incompatible with the present formulation
and interpretation of QSM/QIT. To this end, we concentrate on an
important special class of systems that we call ``strictly
isolated". A system is strictly isolated if and only if (a) it
interacts with no other system in the universe, and (b) its state
is at all times uncorrelated from the state of any other system in
the universe.

The argument that ``real systems can never be strictly isolated and
thus we should dismiss this discussion as useless  at the outset'' is
at once conterproductive, misleading and irrelevant, because the
concept of strictly isolated system is a keystone of the entire
conceptual edifice in physics, particularly indispensable to
structure the principle of causality.  Hence, the strictly isolated
systems must be accepted, at least, as conceivable, in the same way
as we accept within QM that a vector in Hilbert space may represent a
state of a system. Here we take as an essential necessary requirement
that, when applied to a conceivable system and in particular to an
isolated system, the formulation of a physical theory like QSM must
be free of internal conceptual inconsistencies.

In QM the states of a strictly isolated system are in one-to-one
correspondence with the one-dimensional orthogonal projection
operators on the   Hilbert space of the system.  We denote such
projectors by the symbol $P$.  If   $|\psi\rangle$ is an
eigenvector of $P$ such that $P|\psi\rangle = |\psi\rangle$ and
$\langle\psi|\psi\rangle = 1$ then $P = |\psi\rangle\langle\psi|$.
It is well known that differently from classical states, quantum
states are characterized by irreducible intrinsic probabilities.
We give this for granted here, and do not elaborate further on
this point.

Admittedly, the objective of QSM is to deal with situations in
which the state of the   system is not known with certainty. Such
situations are handled, according to von Neumann \cite{Neumann}
(but also to Jaynes \cite{Jaynes1,Jaynes2} within the QIT
approach) by assigning to each of the possible states of the
system an appropriate statistical weight which describes an
``extrinsic" (we use this term to contrast it with ``intrinsic")
uncertainty as to whether that state is the actual state of the
system.  The selection of a rule for a proper assignment of the
statistical weights is not of concern to us here.

To make clear the meaning of the words extrinsic and intrinsic,
consider the following non quantal example. We have two types of
``biased" coins $A$ and $B$ for which ``heads" and ``tails" are
not equally likely. Say that $p_A=1/3$ and $1-p_A=2/3$ are the
intrinsic probabilities of all coins of type $A$, and that
$p_B=2/3$ and $1-p_B=1/3$ those of coins of type $B$. Each time we
need a coin for a new toss, however, we receive it from a slot
machine that first tosses an unbiased
 coin $C$ with intrinsic probabilities $w=1/2$ and
$1-w=1/2$  and, without telling us the outcome, gives us a coin of
type $A$ whenever coin $C$ yields ``head" and a coin of type $B$
whenever $C$ yields ``tail". Alternatively, we pick coins out of a
box where 50\% coins of type $A$ and 50\% coins of type $B$ have been
previously mixed. It is clear that for such a preparation scheme, the
probabilities $w$ and $1-w$ with which we receive (pick up) coins of
type $A$ or of type $B$ have ``nothing to do" with the intrinsic
probabilities $p_A$, $1-p_A$, and $p_B$, $1-p_B$ that characterize
the biased coins we will toss. We therefore say that $w$ and $1-w$
are extrinsic probabilities, that characterize the heterogeneity of
the preparation scheme rather than features of the prepared systems
(the coins). If on each coin we receive we are allowed only a single
toss (projection measurement?), then due to the particular values
($p_A=1/3$, $p_B=2/3$ and $w=1/2$) chosen for this tricky preparation
scheme, we get ``heads" and ``tails" which are equally likely; but if
we are allowed repeated tosses (non-destructive measurements, gentle
measurements, quantum cloning measurements, continuous time
measurements?) then we expect to be able to discover the trick. Thus
it is only under the single-toss constraint that we would not loose
if we base our bets on a description of the preparation scheme that
simply weighs the intrinsic probabilities with the extrinsic ones,
i.e., that would require us to expect ``head" with probability
$p_{\rm head}=wp_A+(1-w)p_B=1/2*1/3+1/2*2/3=1/2$.

For a strictly isolated system, the possible states according to
QM are, in principle, all the one-dimensional projectors $P_i$ on
the Hilbert space $\Hil$ of the system. Let  $\cal P$ denote the
set of all such one-dimensional projectors on $\Hil$. If we are
really interested in characterizing unambiguously a preparation
scheme that yields states in the set $\cal P$ with some
probability density, we should adopt a measure theoretic
description as proposed in Ref.~\citen{thesis}, and define a
``statistical weight measure" $\mu$ satisfying the normalization
condition $\mu({\cal P})=\int_{\cal P} \mu ( d P)=1$ and such that
the expected value of an observable $A$ (which on the base states
is given by $\Tr (PA)$) is given $\langle A\rangle=\int_{\cal P}
\Tr (PA) \mu ( d P)$. As shown in Ref.~\citen{thesis},  this
description would not lead to the kind of ambiguities we are lead
to by adopting the von Neumann description, but it would not lead
to the von Neumann density operator either.

Instead, following the von Neumann recipe, QSM and QIT assign to
each state $P_i$ a statistical weight $w_i$, and characterizes the
extrinsically uncertain situation by a (von Neumann) statistical
 operator $W =\sum_i w_iP_i$, a weighted sum of the
projectors representing the possible states ($W$ is more often
called the density operator and denoted by $\rho$, but we prefer
to reserve this symbol for the state operators we define in the
next section).

The von Neumann  construction is ambiguous, because the same
statistical operator is assigned to represent a variety of different
preparations, with the only exception of homogeneous preparations
({\it proper} preparation in the language of Ref. \cite{Despagnat})
where there is only one possible state $P_\psi$ with statistical
weight 100\% so that $W=W^2=P_\psi$ is ``pure". Given a statistical
operator $W$ (a nonnegative, unit-trace, self-adjoint operator on the
Hilbert space of the system), its decomposition into a weighted sum
of one-dimensional projectors $P_i$ with weights $w_i$ implies that
there is a preparation such that the system is in state $P_i$ with
probability $w_i$. The situation described by $W$ has no extrinsic
uncertainty if and only if $W$ equals one of the $P_i$'s, i.e., if
and only if $W^2 = W = P_i$ (von Neumann's theorem \cite{Neumann}).
Then, QSM reduces to QM and no ambiguities arise.

The problem is that whenever $W$ represents a situation with
extrinsic uncertainty ($W^2 \ne W$) then the decomposition of $W$
into a weighted sum of one-dimensional projectors is not unique.
This is the essence of Schr\"odinger's corollary
\cite{Schroedinger} relevant to this issue (for a mathematical
generalization see Ref.~\citen{Kirkpatrick} and for interpretation
in the framework of non-local effects see e.g.
Ref.~\citen{Mermin}).

For our purposes, notice that every statistical (density) operator
$W$, when restricted to its range $\Ran(W)$, has an inverse that
we denote by $W^{-1}$.  If $W \ne W^2$, then $\Ran(W)$ is at least
two-dimensional, i.e., the rank of $W$ is greater than 1. Let
$P_j=|\psi_j\rangle\langle \psi_j|$ denote the orthogonal
projector onto the one-dimensional subspace of $\Ran(W)$ spanned
by the $j$th eigenvector $|\psi_j\rangle$ of an eigenbasis of the
restriction of $W$ to its range $\Ran(W)$ ($j$ runs from 1 to the
rank of $W$). Then, $W = \sum_j w_jP_j$ where $w_j$ is the $j$-th
eigenvalue, repeated in case of degeneracy. It is noteworthy that
$w_j = [\Tr_{\Ran(W)}(W^{-1}P_j)]^{-1}$. Schr\"odinger's corollary
states that, chosen an arbitrary vector $\alpha_1$ in $\Ran(W)$,
it is always possible to construct a set of vectors
$|\alpha_k\rangle$ ($k$ running from 1 to the rank of $W$,
$\alpha_1$ being the chosen vector) which  span $\Ran(W)$ (but are
not in general
 orthogonal to each other), such that the orthogonal projectors
 $P'_k=|\alpha_k\rangle\langle \alpha_k|$ onto the
 corresponding one-dimensional subspaces of $\Ran(W)$ give rise to the
 alternative resolution of the statistical operator $W = \sum_k
 w'_kP'_k$, with $w'_k =
[\Tr_{\Ran(W)}(W^{-1}P'_k)]^{-1}$.

To fix ideas, consider the example of a qubit with the statistical
operator given by $W=p|1\rangle\langle 1|+(1-p)|0\rangle\langle
0|$ for some given $p$, $0<p<1$. Consistently with Schr\"odinger's
corollary, it is easy to verify that the same $W$ can also be
obtained as a statistical mixture of the two projectors
$|+\rangle\langle +|$ and $|a\rangle\langle a|$ where
$|+\rangle=(|0\rangle +|1\rangle)/\sqrt{2}$, $|a\rangle=(|+\rangle
+ a|-\rangle)/\sqrt{1+a^2}$ (note that $|a\rangle$ and $|+\rangle$
are not orthogonal to each other), $|-\rangle=(|0\rangle-
|1\rangle)/\sqrt{2}$, $a=1/(1-2p)$ and $w=2p(1-p)$ so that
$W=w|+\rangle\langle +|+(1-w)|a\rangle\langle a|$. With $p=1/4$
this is exactly the example given by Park in Ref. \cite{Park1}.

QSM forces on us the following interpretation of Schr\"odinger's
corollary.  The first decomposition of $W$ implies that we may have a
preparation which yields the  system   in state $P_j$ with
probability $w_j$, therefore, the system is for sure in one of the
states in the set $\{P_j\}$. The second decomposition implies that we
may as well have a preparation which yields the  system   in state
$P'_k$ with probability $w'_k$ and,  therefore, the system is for
sure in one of the states in the set $\{P'_k\}$. Because both
decompositions hold true simultaneously, the very rules we adopted to
construct the statistical operator $W$ allow us to conclude that the
state of the system is certainly one in the set $\{P_j\}$, but
concurrently it is also certainly one in the set $\{P'_k\}$. Because
the two sets of states $\{P_j\}$ and $\{P'_k\}$ are different (no
elements in common), this would mean that the system ``is"
simultaneously ``in" two different states, thus contradicting our
starting assumption that a system is always in one definite state
(though perhaps unknown). Little emphasis is gained by noting that,
because the possible different decompositions are not just two but an
infinity, we are forced to conclude that the system is concurrently
in an infinite number of different states! Obviously such conclusion
is   unbearable and perplexing, but it is unavoidable within the
current formulation of QSM/QIT. The reason why we have learnt to live
with this issue -- by simply ignoring it -- is that if we forget
about interpretation and simply use the mathematics, so far we always
got successful results that are in good agreement with experiments.

Also for the coin preparation example discussed above, there are
infinite ways to provide 50\% head and 50\% tail upon a single toss
of a coin chosen randomly out of a mixture of two kinds of biased
coins of opposite bias. If we exclude the possibility of performing
repeated (gentle) measurements on each single coin, than all such
situations are indeed equivalent, and our adopting the weighted sum
of probabilities as a faithful representation is in fact a tacit
acceptance of the impossibility of making repeated measurements. This
limitation amounts to accepting that the extrinsic probabilities
($w$,$1-w$) combine irreducibly with the intrinsic ones
($p_A$,$p_B$), and once this is done there is no way to separate them
again (at least not in a unique way). If these mixed probabilities
are indeed all that we can conceive, then  we must give up the
assumption that each coin has its own possibly unknown, but definite
bias, because otherwise we are lead to a contradiction, for we would
conclude that there is some definite probability that a single coin
has at once two different biases (a monster coin which belongs
concurrently to both the box of, say, 2/3 -- 1/3 biased coins and to
the box of, say, 3/4 -- 1/4 biased coins).

\section{Is there a way out?}

In this section we discuss four main alternatives towards the
resolution of the paradox, that is, if we wish to clear our everyday,
already complicated life from quantum monsters. Indeed, even though
it has been latent for fifty years and it has not impeded major
achievements, the conceptual tension denounced by Schr\"odinger and
Park is untenable, and must be resolved.

Let us therefore restate the three main hinges of QSM which lead to
the logical inconsistency:

\begin{enumerate}
\item  a system is always in a definite, though perhaps unknown,
state; \item states (of strictly isolated systems) are in one-to-one
correspondence with the one-dimensional projectors $P$ on the Hilbert
space $\Hil$  of the system; and   \item statistics of measurement
results from a heterogeneous preparation with extrinsic uncertainty
(probabilities $w_i$) as to which is the actual state of the system
among a set $\{P_i\}$ of possible states is  described by the
statistical operator $W = \sum_i w_i P_i$.
\end{enumerate}
 To
remove the inconsistency, we must reject or modify at least one of
these statements.  But, in doing so, we cannot afford to contradict
any of the   innumerable successes of the present mathematical
formulation of QSM.

A first alternative was discussed by Park \cite{Park1} in his
essay on the nature of quantum states.  If we decide to retain
statements (2) and (3), then we must reject statement (1), i.e.,
we must conclude that the concept of state is ``fraught with
ambiguities and should therefore be avoided."  A system should
never be regarded as being in any physical state. We should
dismiss as unsound all statements of this type: ``Suppose an
electron is in state $\psi$ \dots" Do we need to undertake this
alternative and therefore abandon deliberately  the concept of
state ?  Are we ready to face all the ramifications of this
alternative ?

A second alternative is to retain statements (1) and (2), reject
statement (3) and reformulate the mathematical description of
situations with extrinsic uncertainty in a way not leading to
ambiguities.  To our knowledge,   such a reformulation has never been
considered.  The key defect of the representation by means of
statistical operators is that it mixes irrecoverably two different
types of uncertainties:  the intrinsic uncertainties inherent in the
quantum states and the extrinsic uncertainties introduced by the
statistical description.

In Ref. \citen{thesis}, we have suggested a measure-theoretic
representation that would achieve the desired goal of   keeping the
necessary separation between intrinsic quantal uncertainties and
extrinsic statistical uncertainties. We will elaborate on such
representation elsewhere. Here, we point out that a change in the
mathematical formalism involves the serious risk of contradicting
some of the successes of the present formalism of QSM.  Such
successes are to us sufficient indication that changes in the present
mathematical formalism should be resisted unless the need becomes
incontrovertible.

A third alternative is the QIT approach proposed by Jaynes
\cite{Jaynes1,Jaynes2} and subsequent literature. The paradox is
bypassed (rather than resolved) by introducing an {\it ad-hoc}
``recipe" whereby base states other than eigenstates of the
statistical operator $W$ are to be excluded as unconceivable,
based on the belief that they do not represent ``mutually
exclusive event" \cite{Cubukcu}. We skip here the well-known
details of the QIT {\it ad-hoc} recipe \cite{Jaynes1,Jaynes2} to
obtain the maximal $-\Tr(W\ln W)$ statistical operator $W$ which
should provide the ``best, unbiased description" of the statistics
of measurement results. We need only point out, for the purpose of
our discussion, that such recipe leads to the correct physical
results (i.e., canonical and grand-canonical thermodynamic
equilibrium distributions) only if (1) the experimenter is assumed
to know the value of the energy of the system, not of some other
observable(s); (2) the underlying pure components of the
heterogeneous preparation are ``mutually exclusive" in the sense
that they are the eigenvectors of the Hamiltonian operator of the
system. Then, QIT reduces to equilibrium QSM and expectation
values are successfully computed (from the pragmatic point of
view)  by the formula $\langle A\rangle=\Tr(A W)$ where $W=
\exp(-\beta H)/\Tr[\exp(-\beta H)]$ (or its grand-canonical
equivalent). However, from the conceptual point of view, the two
{\it ad-hoc} conditions just underlined are in clear conflict with
the purely subjective interpretation assumed at the outset in the
QIT approach, for they exclude choices that a truly unbiased
experimenter has no reason to exclude {\it a priori}. In other
words, the fact that such conditions are necessary to represent
the right physics, implies that they represent objective (rather
than subjective) features of physical reality. In particular, they
impose that among the many possible decompositions of the  maximal
$-\Tr(W\ln W)$ statistical operator $W$, which exist by
Schr\"odinger's corollary, the observer is allowed to give a
physical meaning only to the spectral decomposition, thereby being
forced by the recipe to an extremely biased perspective. So, by
ignoring and bypassing the Schr\"odinger-Park conceptual paradox,
the QIT  approach not only does not resolve it, but it opens up
additional conceptual puzzles. For example, what should $W$ be if
the experimenter knows the value of a property other than energy,
or is to describe statistics from a heterogeneous preparation
which is a mixture of pure preparations corresponding to
non-mutually-orthogonal QM states (non-mutually-exclusive events)?
From the application point of view, practitioners in the chemical
physics literature have devised successful modeling and
computational recipes based on constrained maximal entropy
\cite{LaPenna1,LaPenna2} or rate-controlled constrained maximal
entropy \cite{Keck1,Keck2}  in which the energy constraint is
replaced by or complemented with suitably selected other
constraining quantities, e.g., configurational averages
\cite{LaPenna1,LaPenna2} or potentials globally characterizing a
class of slow rate-controlling reaction schemes
\cite{Keck1,Keck2}. But the empirical success of these approaches,
in our view, corroborates the need for  further discussions about
the subjectivity-objectivity conceptual dilemma which remains
unresolved.

A fourth intriguing alternative has been first proposed by
Hatsopoulos and Gyftopoulos \cite{HG1,HG2,HG3,HG4} in 1976. The
idea is to retain statement (1) and modify statement (2) by
adopting and incorporating the mathematics of statement (3) to
describe the true physical states, i.e., the homogeneous
preparations, and at the same time devoiding heterogeneous
preparations (and, therefore, extrinsic statistics) of any
fundamental role. The defining features of the projectors $P$,
which represent the states for a strictly isolated system  in QM,
are: $P^\dagger = P$, $P
> 0$, $\Tr P = 1$, $P^2 = P$.  The defining features of   the
statistical (or density) operators $W$ are $W^\dagger = W$, $W > 0$,
$\Tr W = 1$. Hatsopoulos and Gyftopoulos propose to modify statement
(2) as follows:
\begin{enumerate}
\item[(2')] (HG ansatz) States (of every strictly isolated system) are
in one-to-one correspondence        with the state operators $\rho$
on $\Hil$, where $\rho^\dagger=\rho$, $\rho
> 0$, $\Tr\rho = 1$, without the restriction $\rho^2 = \rho$.
We call these the ``state operators" to emphasize that they play the
same role that in QM is played by the projectors $P$, according to
statement (2) above, i.e., they are associated with the homogeneous
(or pure or proper) preparation schemes.
\end{enumerate}

  Mathematically, state operators $\rho$ have the same
defining features as the   statistical (or density) operators $W$.
But their physical meaning according to statement (2') is sharply
different. A state operator $\rho$ represents a state.  Whatever
uncertainties and probabilities it entails, they are intrinsic in
the state, in the same sense as uncertainties   are intrinsic in a
state described (in QM) by a projector $P =
|\psi\rangle\langle\psi|$. A statistical operator W, instead,
represents (ambiguously) a mixture of intrinsic and extrinsic
uncertainties obtained via a heterogeneous preparation.  In Refs.
\citen{HG1,HG2,HG3,HG4}, all the successful mathematical results
of QSM are re-derived for the state operators $\rho$.  There, it
is shown that statement (2') is non-contradictory to any of the
(mathematical) successes of the present QSM theory, in that region
where  theory is backed by experiment. However it   demands a
serious re-interpretation of such successes because they now
emerge no longer as statistical results (partly intrinsic and
partly extrinsic probabilities), but as non-statistical
consequences (only intrinsic probabilities) of the nature of the
individual states.

In addition, statement (2') implies the existence of a broader
variety of states than conceived of in QM (according to statement
(2)). Strikingly, if we adopt statement (2') with all its
ramifications, those situations in which the state of the system is
not known with certainty stop playing the perplexing central role
that in QSM is necessary to justify the   successful mathematical
results such as canonical and grand canonical equilibrium
distributions. The physical entropy that has been central in so many
discoveries in physics,  would have finally gained its deserved right
to enter the edifice from the front door. It would be  measured by
$-\Boltz\Tr\rho\ln\rho$ and by way of statement (2') and be related
to intrinsic probabilities, differently from the von Neumann measure
$-\Tr W\ln W$ which measures the state of uncertainty determined by
the extrinsic probabilities of a heterogeneous preparation. We would
not be anymore embarrassed by the inevitable need to cast our
explanations of single-atom, single-photon, single-spin heat engines
in terms of entropy, and entropy balances.

 The same observations would be true even in the classical limit
\cite{JMP1}, where the state operators tend to distributions on
phase-space.  In that limit, statement (2') implies a broader variety
of individual classical states than those conceived of in Classical
Mechanics (and described by the Dirac delta distributions on
phase-space). The classical phase-space distributions,   that are
presently interpreted as statistical descriptions of situations with
extrinsic uncertainty, can be readily reinterpreted as
non-statistical   descriptions of individual states with intrinsic
uncertainty. Thus, if we accept this fourth alternative, we must
seriously reinterpret, from a new non-statistical perspective, all
the successes not only of quantum theory but   also of classical
theory.

If we adopt the HG ansatz, the problem of describing statistics of
measurement results from heterogeneous preparations loses the
fundamental role it holds in QSM by virtue of statement (3).
Nevertheless, when necessary, the problem can be unambiguously
addressed as follows \cite{thesis}:
\begin{enumerate}
\item[(3')]  Preparations of  a given system are in one-to-one
correspondence with the normalized measures $ \mu $ that can be
defined on the HG ``quantal state domain of the system'', $
\mathcal{R} $, i.e., the set of all possible state operators
$\rho$ on $\Hil$ defined according to statement (2') [the
normalization condition is $ \mu ( \mathcal{R} ) =
\int_{\mathcal{R}} \mu ( d \rho ) = 1$]. We call each such measure
$ \mu $ a ``statistical-weight measure over the quantal
phase-domain of the system''. Statistics of measurement results
from a heterogeneous preparation with extrinsic uncertainty
(probabilities $w_i$) as to which is the actual state of the
system among a discrete set $\{\rho_i\}$ of possible states is
described by the statistical-weight measure  $\mu = \sum_i w_i
\mu_{\rho_i}$ where $\mu_{\rho_i}$ is the Dirac measure
``centered" at state $\rho_i$.\cite{footnote1}
\end{enumerate}
The discussion of such description, first introduced in Ref.
\cite{thesis}, is not essential here and will therefore be
presented elsewhere (recently, some useful mathematical results
have been developed along these lines, but in another context, in
Refs. \cite{Zapatrin1,Zapatrin2}). For the present purpose it
suffices to say that the Dirac measures are the only irreducible
measures that can be defined over $ \mathcal{R} $ \cite{thesis}.
In fact, any other measure can be decomposed  {\it in a unique
way} into a ``sum" of Dirac measures and is therefore reducible.
The physical meaning of the uniqueness of the ``spectral"
resolution of any measure into its component Dirac measures is
that the statistical descriptor $\mu$ associated with any
preparation is complete and unambiguous, because its unique
``spectral" resolution identifies unambiguously every component
homogeneous preparation through the support of the corresponding
Dirac measure, as well as the respective statistical weight. As a
result, this mathematical description of heterogeneous
preparations does not lead to  the Schr\"odinger-Park paradox and
hence the concept of state is saved.\cite{footnote2}

\section{Concluding remarks}

 In conclusion, the Hatsopoulos-Gyftopoulos ansatz, proposed thirty
 years ago in Refs.  \citen{HG1,HG2,HG3,HG4} and follow up theory
 \cite{NATO,Cimento1,Cimento2,Nature,Onsager,ArXiv1,MPLA,PRE},
 not only resolves the Schr\"odinger-Park paradox
 without rejecting the concept of state (a keystone of scientific thinking),
 but forces us to  re-examine
  the physical nature of
the individual states (quantum and classical), and finally gains for
thermodynamics and in particular the second law a truly fundamental
role, the prize it deserves not only for having never failed in the
past 180 years since its discovery by Carnot, but also for having
been and still being a perpetual source of reliable advise as to how
things work in Nature.

In this paper, we  restate a seldom recognized conceptual
inconsistency which is unavoidable within the present formulation
of QSM/QIT and discuss briefly logical alternatives towards its
resolution. Together with Schr\"odinger \cite{Schroedinger} who
first surfaced the paradox and Park \cite{Park1,Park2} who first
magistrally explained the
  incontrovertible tension it introduces around the fundamental concept of state of a system,
  we maintain that this fundamental difficulty is by itself a
sufficient reason to go beyond QSM/QIT, for we must resolves the
``essential tension" which has sapped the conceptual foundations of
the present formulation of quantum theory for almost eighty years.

  We argue that rather than adopting the drastic way out provokingly prospected by Park,
  namely, that we should
   reject as unsound the very concept of
state of a system (as we basically do every day by simply ignoring
the paradox), we may alternatively remove the paradox by rejecting
the present statistical interpretation of QSM/QIT without
nevertheless rejecting the successes of its mathematical formalism.
The latter resolution is satisfactory both conceptually and
mathematically, but  requires that the physical meaning of the
formalism be reinterpreted with care and detail. Facing the situation
sounds perhaps uncomfortable because there seems to be no harmless
way out, but if we adopt the Hatsopoulos-Gyftopoulos fundamental
ansatz (of existence of a broader kinematics) the change will be at
first mainly conceptual, so that practitioners who happily get
results everyday out of QSM would basically maintain the {\it status
quo}, because we would
 maintain the same mathematics both for the time-independent state
operators that give us the canonical and grand-canonical description
of thermodynamics equilibrium states, and for the time-dependent
evolution of the idempotent density operators ($\rho^2=\rho$), i.e.,
the states of ordinary QM, which keep evolving unitarily. On the
other hand, if the ansatz is right, new physics is likely to  emerge,
for it would imply that beyond the the states of ordinary QM, there
are states (``true" states, obtained from preparations that are
``homogeneous" in the sense of von Neumann \cite{Neumann}) that even
for an isolated and uncorrelated single degree of freedom ``have
physical entropy" ($-\Boltz\Tr\rho\ln\rho$) and require a
non-idempotent state operator ($\rho^2\ne\rho$) for their
description, and therefore exhibit even at the microscopic level the
limitations imposed by the second law,

In addition, if we  adopt as a further ansatz that the time
evolution of these non-ordinary-QM states (the non-idempotent
ones) obeys the nonlinear equation of motion developed by the
present author
\cite{NATO,Cimento1,Cimento2,Nature,Onsager,MPLA,PRE}, then in
most cases they do not evolve unitarily but follow a path that
results from the competition of the Hamiltonian unitary propagator
and a new internal-redistribution propagator that ``pulls" the
state operator $\rho$ in the direction of steepest entropy ascent
(maximal entropy generation) until it reaches a (partially)
canonical form (or grand canonical, depending on the system). Full
details can be found in Refs. \citen{Cimento2,ArXiv1}.

 The proposed
resolution definitely goes beyond QM, and turns out to be in line
with Schr\"odinger's prescient
 conclusion of his 1936 article \cite{Schroedinger} when he writes:
``My point is, that in a domain which the present theory does not
cover, there is room for new assumptions without necessarily
contradicting the theory in that region where it is backed by
experiment."

\section{Acknowledgements}
The author is indebted to Lorenzo Maccone for a helpful
discussion.


\begin{thebibliography}{33}
\bibitem{footnote1}Among the measures that can be defined over  $
\mathcal{R} $, with every state operator $\rho_o$ in $ \mathcal{R}
$ we can associate a Dirac measure defined as follows
\cite{Fano,thesis}. Let $ E $ denote any subset of $ \mathcal{R}
$, then $\mu_{\rho_o} ( E ) = 1$ if $ \rho_o \in E $ and
$\mu_{\rho_o} ( E ) =   0$ if $ \rho_o \notin E $. The {\it
support} of a measure is the subset of the domain $ \mathcal{R} $
for which the measure is nonzero. Clearly the Dirac measure
$\mu_{\rho_i}$ has a single-point support coinciding with the
state operator $\rho_i$.
\bibitem{footnote2}Notice that within standard QM, where states of
strictly isolated systems are one-to-one with the unit-norm
vectors $\psi$ in the Hilbert space $\Hil$ of the system, a
natural and unambiguous description of the statistics from a
heterogeneous preparation can be obtained by using (instead of the
von Neumann statistical operator $W$) the normalized
statistical-weight measures $\mu$ defined on the set of all
possible unit-norm (pure state) vectors $\psi$ in $\Hil$. The
uniqueness of the decomposition $\mu = \sum_i w_i \mu_{\psi_i}$
into the component Dirac measures (in the discrete case) or its
continuous version (Refs. \cite{Zapatrin1,Zapatrin2}) would not
give rise to the Schr\"odinger-Park paradox. However, as we will
show elsewhere, the maximal-statistical-uncertainty measures that
would correspond to thermodynamic equilibrium according to
standard QSM/QIT reasoning, would differ in general from the
canonical and grand canonical distributions!
\end{thebibliography}
\end{document}